\documentstyle[12pt]{article}

\begin{document}
\renewcommand{\thefootnote}{\fnsymbol{footnote}}
\newcommand{\be}{\begin{equation}}\newcommand{\ee}{\end{equation}}
\newcommand{\bea}{\begin{eqnarray}}\newcommand{\eea}{\end{eqnarray}}
\newcommand{\nn}{\nonumber}\newcommand{\p}[1]{(\ref{#1})}
\newpage
\pagestyle{empty}
\setcounter{page}{0}

\newcommand{\norm}[1]{{\protect\normalsize{#1}}}
\newcommand{\LAP}
{{\small E}\norm{N}{\large S}{\Large L}{\large A}\norm{P}{\small P}}
\newcommand{\sLAP}{{\scriptsize E}{\footnotesize{N}}{\small S}{\norm
L}$
${\small A}{\footnotesize{P}}{\scriptsize P}}
\begin{minipage}{5.2cm}
\begin{center}
{\bf G{\sc\bf roupe d'} A{\sc\bf nnecy}\\
\ \\
Laboratoire d'Annecy-le-Vieux de Physique des Particules}
\end{center}
\end{minipage}
\hfill
\hfill
\begin{minipage}{4.2cm}
\begin{center}
{\bf G{\sc\bf roupe de} L{\sc\bf yon}\\
\ \\
Ecole Normale Sup\'erieure de Lyon}
\end{center}
\end{minipage}
\centerline{\rule{12cm}{.42mm}}

\vspace{10mm}

\begin{center}
{\bf \Large{On non-linear superfield versions of the
vector-tensor multiplet}}
\\[1cm]

\vspace{5mm}

{\large E. Ivanov$^{a}$} \\
{\em  Bogoliubov Laboratory of Theoretical Physics,
Joint Institute for Nuclear Research\\ 141980 Dubna, Russia}\\
\bigskip
{\large and} \\
\bigskip
{\large E.  Sokatchev$^{b}$} \\
{\em Laboratoire de Physique Th\'eorique }\LAP\footnote{URA 14-36 du
CNRS, associ\'ee \`a l'Ecole Normale Sup\'erieure de Lyon et \`a
l'Universit\'e de Savoie

\noindent
$^a$ e-mail address: eivanov@thsun1.jinr.dubna.su \\
\noindent
$^b$ e-mail address: sokatche@lapp.in2p3.fr \\
\noindent
}\\
Groupe d'Annecy: LAPP, Chemin de Bellevue BP 110\\ F-74941
Annecy-le-Vieux Cedex, France

\end{center}
\vspace{10mm}

\centerline{ {\bf Abstract}}
\vskip5mm
\indent
We propose a harmonic superspace description of the non-linear vector-tensor
$N=2$ multiplet. We show that there exist two inequivalent version: the old one
in which one of the vectors is the field-strength of a gauge two-form, and a
new one in which this vector satisfies a non-linear constraint and cannot be
expressed in terms of a potential. In this the new version resembles the
non-linear $N=2$ multiplet. We perform the dualization of both non-linear
versions in terms of a vector gauge multiplet and discuss the resulting
holomorphic potentials. Finally, we couple the non-linear vector-tensor
multiplet to an abelian background gauge multiplet.

\vfill
\rightline{hep-th/9711038}
\rightline{\LAP-A-667/97}
\rightline{JINR E2-97-336}
\rightline{ November 1997}

\newpage
\pagestyle{plain}
\renewcommand{\thefootnote}{\arabic{footnote}}
\setcounter{footnote}{0}

\newpage\setcounter{page}1

\section{Introduction}
Recently, there was a revival of interest in the $N=2$ vector-tensor (VT)
multiplet \cite{Kelly}, mainly due to the fact that it describes the
axion/dilaton complex in heterotic $N=2$ four-dimensional supersymmetric
string vacua \cite{WKLL}. The VT multiplet is a variant representation of
the $N=2$ vector multiplet, such that one of the physical scalars of the
latter is traded for a gauge antisymmetric tensor (notoph) off shell. The
known off-shell formulation of the VT multiplet ($8 + 8$ components)
necessarily implies the presence of a central charge in the $N=2$
superalgebra. It is real and acts on the component fields in a highly
non-trivial way. As was observed in \cite{dW1},\cite{dW2}, there exist at
least two different versions of the VT multiplet. Their basic difference is
in the coupling of the tensor and vector gauge fields: in the so-called
``non-linear'' version the tensor field couples to the Chern-Simons (CS)
form of the vector one, while no such CS coupling is present in the
``linear'' version. $N=2$ supersymmetry is realized in these two cases in
essentially different ways: non-linearly in the first case and linearly in
the second one. The two versions also radically differ in what concerns
couplings to background $N=2$ vector multiplets and $N=2$ supergravity
\cite{dW2,dW3}. Note that the central charge transformations can be global
or local. In the latter case an extra vector multiplet gauging the central
charge should be introduced from the beginning. When coupling the VT
multiplet to supergravity, the central charge is necessarily gauged.

An exhaustive component analysis of the two versions of the VT multiplet
together with their couplings to background vector multiplets was given in
\cite{dW1}, \cite{dW2}, assuming that the central charge transformations
are local. As for the superfield formulations of the VT multiplet (which
are most natural when dealing with off-shell supermultiplets), until
recently they were constructed only for the linear version, both in the
free case and in the presence of couplings to background vector multiplets
\cite{Ovr1}, \cite{Ovr2}, \cite{Grimm}, \cite{DK1}. There exist
formulations in the standard \cite{Ovr1} - \cite{Grimm} as well as in
harmonic \cite{DK1} $N=2$ superspaces\footnote{When this work was near
completion, we became aware of the parallel work \cite{DK2} where a
harmonic superspace formulation of the non-linear version of the VT
multiplet was given (at the level of rigid central charge and without
considering CS couplings to extra vector multiplets).}.

Our purpose in this letter is to give such a formulation for the non-linear
version, both for the pure VT multiplet and for the case when CS couplings
to the background vector multiplets are switched on. We make use of the
harmonic superspace (HSS) approach as most adequate to $N=2$ supersymmetry
and demonstrate the existence of {\it two inequivalent non-linear versions}
of the VT multiplet. The first one is just the version discovered in
\cite{dW1,dW2}, while the second is essentially new: it cannot be reduced
to the ``old'' one by any field redefinition. Its most characteristic
feature is the modification of the r.h.s. of the Bianchi identity for the
three-form (the field strength of the tensor gauge field) by terms
quadratic in the latter and in the auxiliary fields. As a result, the
Bianchi identity has no local solution in terms of a tensor gauge potential
(note that one of the primary assumptions of \cite{dW1,dW2} is the
existence of such a potential). We show that the bosonic action of this new
version of the VT multiplet vanishes as a consequence of the modified
Bianchi identity for the three-form. Nevertheless, a non-trivial action is
obtained upon dualization, i.e. after implementing this identity in the
action with a scalar Lagrange multiplier. In this aspect, the situation is
quite similar to the case of non-linear $N=2$ multiplet \cite{nonl}. The
dual action exhibits all the features of special K\"ahler geometry
typical of the actions of $N=2$ vector multiplets and is fully
specified by a non-polynomial holomorphic potential. We propose a
manifestly supersymmetric version of the dualization procedure.

Here we restrict our study to the rigid case, postponing the discussion of
the gauged central charge and, more generally, $N=2$ supergravity to a
future publication. Also, when discussing the superfield CS couplings of
the non-linear version of the VT multiplet to background vector multiplets,
for the sake of simplicity we consider one abelian multiplet coupled to the
``old'' non-linear version of the VT multiplet. The generalization to an
arbitrary number of background vector multiplets, both abelian and
non-abelian, as well as to the case of the ``new'' non-linear version is
rather straightforward and will be presented elsewhere.

\section{Preliminaries}
Let us first briefly recall some facts about the HSS description of the
linear version of the VT multiplet, to a large extent following ref.
\cite{DK1}. We assume the reader to be familiar with the basic concepts of
the harmonic superspace approach to $N=2$ supersymmetry; otherwise, we
invite him to consult the original papers \cite{{GIKOS},{GIKOS1}} or the
(yet) unpublished review \cite{HSS}.

The basic object of such a description is the real harmonic superfield
$$
L = L(x^{\alpha\dot\alpha}, x^5, \theta^{\pm \alpha},
\bar \theta^{\pm \dot\alpha}, u^{\pm i})
$$
subject to the constraints
\bea
(D^{+})^2 L &=& (\bar D^{+})^2 L = 0\;, \label{lin1} \\
D^+_\alpha \bar D^+_{\dot\alpha} L &=& 0, \label{lin2} \\
D^{++}L &=& 0. \label{harm1}
\eea
Here
\bea
u^{+i}u^-_i &=& 1, \quad \theta^{\pm\alpha} = \theta^{\alpha i}u^{\pm}_{i},
\quad \bar\theta^{\pm \dot\alpha} = \bar \theta^{\dot \alpha i}u^{\pm}_{i},
\quad
D^{+}_\alpha = \frac{\partial}{\partial \theta^{-\alpha}} \equiv
\partial_{-\alpha}, \quad
\bar D^{+}_{\dot\alpha} =
\frac{\partial}{\partial \bar \theta^{-\dot\alpha}} \equiv
\partial_{-\dot\alpha}, \nn \\
\quad D^{++} &=& u^{+i}\frac{\partial}{\partial u^{-i}} -
2i \theta^{+\alpha}\bar\theta^{+\dot\alpha}\partial_{\alpha\dot\alpha}
+ i((\theta^+)^2 - (\bar \theta^+)^2)\partial_5 + \theta^{+\alpha}
\partial_{-\alpha} + \bar \theta^{+\dot\alpha}\bar\partial_{-\dot\alpha}
\label{++}
\eea
are the basic quantities of the central-charge extended HSS in the analytic
basis. This basis is chosen so that the covariant spinor derivatives
$D^+_\alpha, \bar D^+_{\dot\alpha}$ are ``short'' and the coordinate
sets
\be
\zeta^{5} \equiv
\{x^{\alpha\dot\alpha}, x^5, \theta^{+\alpha}, \bar \theta^{+\dot\alpha},
u^{\pm i} \} \label{anal5}
\ee
and
\be
\zeta \equiv
\{ x^{\alpha\dot\alpha}, \theta^{+\alpha}, \bar \theta^{+\dot\alpha} ,
u^{\pm i} \} \label{anal}
\ee
are closed under the $N=2$ supersymmetry transformations. They are called
analytic subspaces. The harmonic derivative $D^{++}$ commutes with
$D^+_{\alpha}, \bar D^+_{\dot\alpha}$, and so it preserves analyticity. In
what follows we will always use the analytic basis in $N=2$ HSS. Note that
the subspaces \p{anal5}, \p{anal} are real, i.e. closed under some
generalized conjugation. Our conventions are those of ref. \cite{GIKOS}.

The set of constraints \p{lin1} - \p{harm1} reduces the infinite component
content of $L$ to that of the off-shell linear VT multiplet when formulated
via the field strengths of the notoph and vector gauge potentials
\footnote{At present it is unclear, even at this simplest linear level,
what could be (if existing!) the HSS description of the VT multiplet in
terms of superfield potentials.}.

It is convenient to represent $L$ by its analytic components, i.e. by the
functions on the subspace $\zeta^5$ which appear in the decomposition of
$L$ in powers of $\theta^{-\alpha}, \bar\theta^{-\dot\alpha}$. The
constraints \p{lin1}, \p{lin2} imply
\be
L = l(\zeta^5) + \theta^-f^+(\zeta^5) + \bar\theta^-\bar f^+(\zeta^5)
\label{dec1}
\ee
(here and in what follows the spinor indices are contracted according to
the rule ${}^\alpha{}_\alpha, {}_{\dot\alpha}{}^{\dot\alpha}$),
while \p{harm1} leads to the two harmonic constraints
\bea
&& D^{++}l +\theta^+ f^+ + \bar\theta^+\bar f^+ = 0~,  \label{harm2a} \\
&& D^{++}f^+ = D^{++}\bar f^+ = 0~.   \label{harm2b}
\eea
Thus an equivalent description of the VT multiplet is given
in terms of the analytic scalar and spinor superfunctions $l$, $f^+_\alpha $,
$\bar f^+_{\dot\alpha}$ subject to
the harmonic constraints \p{harm2a}, \p{harm2b}. Note that $f^+_\alpha$
is transformed under $N=2$ supersymmetry as a
standard analytic superfield while $l$ has unusual transformation
properties:
$$
\delta f^+_\alpha = 0~, \quad
\delta l = -\epsilon^iu^-_if^+ - \bar \epsilon^iu^-_i\bar f^+
$$
where $\epsilon^i_\alpha, \bar\epsilon^i_{\dot\alpha}$ are
infinitesimal transformation parameters.

Eqs. \p{harm2a}, \p{harm2b} fully determine the action of the central charge
generator ${\partial/\partial x^5}$ on the component fields in $l$, $f^+$.
In what follows it will be more convenient to define its
action directly on the analytic quantities $l$, $f^+$.
This can be done using the following trick. As a consequence of the
harmonic condition \p{harm1} we have
\be
D^{--}L = 0~, \label{harm1min}
\ee
where $D^{--}$ is the harmonic derivative conjugate (in the usual sense) to
$D^{++}$
\be
D^{--} = u^{-i}\frac{\partial}{\partial u^{+i}} -
2i \theta^{-\alpha}\bar\theta^{-\dot\alpha}\partial_{\alpha\dot\alpha}
+ i((\theta^-)^2 - (\bar \theta^-)^2)\partial_5 +
\theta^{-\alpha}\partial_{+\alpha} +
\bar \theta^{-\dot\alpha}\bar\partial_{+\dot\alpha}   \label{Dminmin}
\ee
(it does not preserve analyticity!).
Together with $D^{++}$ they form the $SU(2)$ algebra of harmonic
derivatives:
\be
[D^{++}, D^{--}] = D^0~,\quad [D^0, D^{\pm\pm}] = \pm 2 D^{\pm\pm}~,
\ee
where $D^0$ is the operator counting the harmonic $U(1)$ charge ($D^0 l =
0~, \quad D^0f^+ = f^+$). Substituting
\p{dec1} into \p{harm1min} and equating to zero the coefficients in front
of the various powers of $\theta^-$, $\bar\theta^-$, we find
the set of constraints:
\bea
&&\partial^{--}l = 0~, \label{1} \\
&&\partial_{+\alpha}l + \partial^{--}f^+_\alpha = 0~, \quad
\partial_{+\dot\alpha}l - \partial^{--}\bar f^+_{\dot\alpha} = 0~,
\label{2} \\
&&\partial_{\alpha\dot\beta}l +{i\over 2}
(\partial_{+\alpha}\bar f^+_{\dot\alpha} + \partial_{+\dot\alpha}f^+_\alpha)
= 0~, \label{3} \\
&& \partial_5 l -{i\over 2} \partial_{+\alpha}f^{+\alpha} = 0~,
\quad \partial_5 l +{i\over 2} \partial_{+\dot \alpha}\bar f^{+\dot \alpha}
= 0~, \label{4} \\
&& \partial_5 f^+_\alpha - \partial_{\alpha\dot\alpha}\bar f^{+\dot\alpha}
= 0~, \quad \partial_5 \bar f^+_{\dot\alpha} +
\partial_{\alpha\dot\alpha} f^{+\alpha} = 0~, \label{5}
\eea
where $\partial^{--}=u^{-i}{\partial}/{\partial u^{+i}}$.
An important corollary of eqs. \p{4} is the reality condition
\be
\partial_{+\alpha}f^{+\alpha} + \partial_{+\dot\alpha}\bar f^{+\dot\alpha} =
0. \label{real}
\ee
Introducing
\be
D^{-}_\alpha = [D^{+}_\alpha, D^{--}]~, \quad
\bar D^{-}_{\dot\alpha} = [\bar D^{+}_{\dot\alpha}, D^{--}]~,  \label{Dmin}
\ee
it is easy to show that another form of \p{real} is
\be
D^{-\alpha}f^+_\alpha = \bar D^-_{\dot\alpha}\bar f^{+\dot\alpha},
\ee
which is just the reality condition of ref. \cite{DK1}. In our approach it
is clear that this condition is a direct consequence of the choice
of a real central charge (had we chosen $x^5$ to be complex, the two eqs.
\p{4} would be independent and no relation of the sort \p{real} would
arise).

Combining relations \p{1} - \p{5} with eqs. \p{harm2a}, \p{harm2b},
it is easy to find out the irreducible field content of $l$, $f^+$
and to show that it exactly coincides with that of the linear version
of the VT multiplet:
\bea
&& l| = \phi(x,x^5)~, \quad \partial_5 \phi \equiv G(x,x^5)~,
\quad f^+_\alpha | = f^i_\alpha (x,x^5) u^+_i, \nn \\
&& \partial_{+\beta} f^+_\alpha | = F_{(\beta \alpha)}(x,x^5) +
i \epsilon_{\beta\alpha} G(x,x^5)\;, \quad \partial_{+\dot\alpha}
f^+_\beta |
= h_{\beta\dot\alpha}(x,x^5) + i \partial_{\beta\dot\alpha} \phi(x,x^5)
\label{comp}
\eea
where $|$ means restriction to the lowest component of a given superfield.
After simple algebraic manipulations involving the above constraints, all
other components, including those obtained by acting on \p{comp} with
$\partial_5$, are expressed as $x$-derivatives of the basic quantities
\p{comp}. For instance,
\be
\partial_5 G = {1\over 2}\partial^{\alpha\dot\alpha}
\partial_{\alpha\dot\alpha}\phi\;.
\ee
The Bianchi identities for $F_{(\alpha\beta)}$, $h_{\alpha\dot\beta}$
also directly follow from the constraints. For instance, acting by
$\partial_5$ on the reality condition \p{real} and on eq. \p{3}, and
making use of eqs. \p{5} afterwards, one gets, respectively,
\be
\partial \cdot h = 0~, \label{notofbia}
\ee
and
\be
\partial_{\alpha\dot\beta}\bar F_{(\dot\alpha}^{\;\;\dot\beta)} -
\partial_{\beta\dot\alpha} F_{(\alpha}^{\;\;\beta)} = 0~, \label{vectbia}
\ee
which are the Bianchi identities for the notoph and vector gauge field
strengths.

In ref. \cite{DK1} there was proposed a nice general recipe of constructing
HSS actions for supermultiplets with a non-trivial realization of the
central charge, such that they are still given by integrals over the
standard analytic subspace \p{anal} containing no $x^5$
coordinate\footnote{Note that a similar HSS action is used to describe the
massive central-charged hypermultiplet \cite{HSS}}. The action is given
by the general formula
\be
S = \int d\zeta^{-4} i[(\theta^+)^2 - (\bar\theta^+)^2] {\cal L}^{++}
\label{action}
\ee
where $d\zeta^{-4}\equiv dud^4xd^4\theta^+$. The real Lagrangian
density ${\cal L}^{++}$ should be:

(i) analytic:
\be
D^+_\alpha {\cal L}^{++} = \bar D^+_{\dot\alpha} {\cal L}^{++} = 0 ~.
\label{analytL}
\ee

(ii) harmonically ``short'':
\be
D^{++}{\cal L}^{++} = 0~. \label{harmL}
\ee

The second condition immediately leads to the important property that the
$x^5$ derivative of the integrand in \p{action} is a total $x$ and $u$
-derivative (recall (\ref{++})) and so disappears upon integration. As a
result the action \p{action} does not depend on $x^5$ or, to put it
differently, is invariant under central charge transformations. The $N=2$
supersymmetry of \p{action} is not manifest, but can be easily
checked (see \cite{DK1}).

In the case under consideration two Lagrangian densities of this sort
exist \cite{DK1}:
\bea
&& (a)\; {\cal L}^{++}_1  = i(D^+LD^+L - \bar D^+L\bar D^+L) =
i(f^+f^+ - \bar f^+\bar f^+)~, \label{act1} \\
&& (b) \; {\cal L}^{++}_2  = (D^+LD^+L + \bar D^+L\bar D^+L) =
(f^+f^+ + \bar f^+\bar f^+)~. \label{act2}
\eea
The first density gives the free action of the linear VT multiplet. The
second one is a total $x$-derivative, i.e. gives a topological invariant.
Both of them, as well as the defining constraints \p{lin1} -
\p{harm1}, can be generalized to include CS couplings to external $N=2$
vector gauge multiplets. These extensions were given in \cite{DK1}. We will
return to this point in Section 4.

\setcounter{equation}{0}
\section{Non-linear VT multiplets}

As was already mentioned, a characteristic feature of the non-linear version
of the VT multiplet discovered in  \cite{dW1,dW2} is the presence of
 CS coupling-induced terms
of the vector gauge field in the Bianchi identity for the notoph gauge field
strength. A simple analysis shows that a minimal way to obtain such terms in
the HSS description is to modify the linear VT multiplet
constraints as follows
\bea
(D^{+})^2 L &=& \alpha (L) D^+LD^+L +
\beta (L) \bar D^+L\bar D^+L \label{nonlin1} \\
D^+_\alpha \bar D^+_{\dot\alpha} L &=& \gamma (L) D^+_\alpha L
\bar D^+_{\dot\alpha} L, \label{nonlin2} \\
D^{++}L &=& 0~, \label{harm2}
\eea
with $\alpha (L)$, $\beta(L)$ being complex and $\gamma (L) = \bar \gamma
(L)$ real functions of $L$, arbitrary for the moment. Note that
\p{nonlin1}, \p{nonlin2} provide the most general deformation of
the linear constraints \p{lin1}, \p{lin2} consistent with the
preservation of the harmonic $U(1)$ charge and the harmonic condition
\p{harm2}. It is worth mentioning that in principle the latter can also be
deformed by adding appropriate bilinears of $D^+_\alpha$,
$\bar D^+_{\dot\alpha}$ into its r.h.s. We do not consider such
non-minimal possibilities here.

The constraints \p{nonlin1}, \p{nonlin2} should satisfy the evident
self-consistency  conditions
\be
D^+_\alpha (D^+)^2 L = 0~, \quad \bar D^+_{\dot\alpha} (D^+)^2 L =
D^{+\alpha} (D^{+}_{\alpha} \bar D^+_{\dot\alpha} L)~, \label{selfcon}
\ee
which amount to the following set of differential equations for
the coefficients
\bea
(\gamma - \alpha)' &=& (\alpha - \gamma)\gamma - \beta\bar\beta
\label{sc1} \\
\beta' &=& (\alpha - 2\gamma)\beta~.  \label{sc2}
\eea
Thus we have four real differential equations for five real functions.
However, we are actually dealing with {\it four} unknowns due to
the reparametrization freedom
\be
L \rightarrow \tilde L~, \quad L = L(\tilde L) \label{reparam}
\ee
in \p{sc1}, \p{sc2}. Under such reparametrizations the coefficients
transform as follows:
\be
\alpha \rightarrow \tilde \alpha = L' \alpha - (\mbox{ln}\; L')'~,
\quad \beta \rightarrow \tilde \beta = L' \beta~,
\quad \gamma \rightarrow \tilde \gamma = L'\gamma - (\mbox{ln}\;L')'~.
\label{rep1}
\ee

We can choose different gauges with respect to \p{rep1} in order to simplify
the set \p{sc1}, \p{sc2}. A very convenient gauge amounts to choosing
\be
\gamma = 0   \label{gauge2}
\ee
which implies
\be
\alpha' = \beta\bar\beta~, \quad \beta' = \alpha \beta \label{eqs2}~.
\ee
In this gauge the constraints (\ref{nonlin1})-(\ref{nonlin2})
become simpler:
\be
(D^+)^2   L =
\alpha D^+  L D^+  L + \beta \bar D^+  L \bar D^+  L ~,
\qquad
D^+_\alpha \bar D^+_{\dot\alpha}   L = 0~.  \label{solgau2}
\ee
The main advantage of the constraints in the form \p{solgau2} is that there
appear no mixed terms in the $\theta^-_\alpha, \bar \theta^-_{\dot\alpha}$
expansion of $L$. Indeed, the solution to the second of eqs. \p{solgau2}
is (cf (\ref{dec1}) in the linear case):
\be
L = l + \theta^-f^+ + \bar\theta^-\bar f^+
-{1\over 4} (\theta^-)^2[\alpha(f^+)^2+\beta(\bar f^+)^2]
-{1\over 4} (\bar\theta^-)^2[\bar\alpha(\bar f^+)^2+\bar\beta(f^+)^2] ~.
\label{decomp}
\ee

It is easy to find the general solution to the equations (\ref{eqs2}), but
before doing this, we point out that additional restrictions on the
coefficient functions $\alpha,\beta$
come from the harmonic condition \p{harm1min}.
Applying the reasoning which lead to eqs. \p{1} - \p{5}, one finds the
analogs of the latter for the non-linear case. Eqs. \p{1},
\p{2} preserve their form, while those from \p{3} on are
modified by non-linear terms:
\bea
&& \partial^{--}l = 0~,  \label{12} \\
&& \partial^{--}f^+_\alpha + \partial_{+\alpha}l = 0~, \quad
\partial^{--}\bar f^+_{\dot\alpha}  - \partial_{+\dot\alpha}l = 0~,
\label{22} \\
&& \partial_{\alpha\dot\alpha} l +{i\over 2}\;(\partial_{+\dot\alpha}
f^+_\alpha  + \partial_{+\alpha}\bar f^+_{\dot\alpha}) = 0~, \label{32} \\
&& \partial_5 l -{i\over 2}(\partial_{+\alpha}f^{+\alpha} - \alpha \;
\partial^{--}f^+ f^+ - \beta \; \partial^{--}\bar f^+ \bar f^+) = 0
\quad \mbox{and c.c.} \label{42} \\
&& \partial_5 f^+_\alpha - \partial_{\alpha \dot\beta} \bar f^{+\dot\beta}
+{i\over 2}\; (\alpha \;\partial_{+\alpha} f^+ f^+  + \beta \;
\partial_{+\alpha} \bar f^+ \bar f^+)  \nn \\
&&+ \;{i\over 4} \;\alpha\beta \;
\partial^{--}f^+_\alpha\; [(f^+)^2 + (\bar f^+)^2] = 0 \quad
\mbox{and c.c.} \label{52}
\eea
A new phenomenon in the non-linear case is the appearance of
a new self-consistency condition as a result of equating
to zero the coefficient of the monomial $(\theta^-)^2(\bar\theta^-)^2$ in
\p{harm1min}. It reads
\be
\partial_5[(\alpha - \bar \beta) (f^+)^2 +
(\beta - \bar \alpha) (\bar f^+)^2] = 0\:. \label{selfcon5}
\ee
Working out the derivative $\partial_5$ and expressing $\partial_5 l,\partial_5
f^+_\alpha,\partial_5 \bar f^+_{\dot\alpha}$ from eqs. (\ref{42}), (\ref{52}),
we see that there appear unacceptable algebraic constraints on the fermions
$f^+$, unless we demand
\be\label{realit}
\alpha=\bar\beta~.
\ee
This new constraint, together with (\ref{eqs2}), imply
\be
\alpha'=\alpha\bar\alpha~.\label{simpler}
\ee
Putting $\alpha=a+ib$ in (\ref{simpler}), we find
 $$
b=\mbox{const}
 $$
and
\be
a'=a^2+b^2~. \label{final}
\ee
The solution to the differential equation \p{final} depends on the
value of the constant $b$. If $b\neq 0$, one obtains
\be \label{neq}
\alpha = b\;[\tan(L+c)+i]~,
\ee
where $c$ is a new integration constant; if $b=0$, the solution is
\be \label{equ}
\alpha =-{1\over L+c}~.
\ee
Note that after choosing the gauge (\ref{gauge2}) we still have the freedom
of global rescalings and shifts of $L$. Using this, we can fix the
constants $b,c$ in (\ref{neq}) or (\ref{equ}), for example, $b=1$, $c=0$.
Thus, in the gauge (\ref{gauge2}) we obtain two distinct solutions:
\be
{\rm (i)} \; \alpha = \tan L + i; \quad {\rm (ii)}\; \alpha = -
{1\over L} ~.
\label{sol12}
\ee
They give rise to {\sl two inequivalent versions} of the non-linear VT
multiplet (remember that we have already exhausted the freedom of
redefinition of $L$).

The principle difference between these two versions is in the following. It
is easy to deduce the analogs of the Bianchi identities \p{notofbia},
\p{vectbia} for both non-linear versions. Eq. \p{vectbia} does not change,
implying that $F_{(\alpha\beta)}$, $F_{(\dot\alpha\dot\beta)}$ are still
expressed in the standard way through the vector gauge potential. At the
same time, the identity \p{notofbia} is drastically modified:
\be
\partial \cdot h + {i\over 4}\; (\alpha \;F^2 -
\bar \alpha \; \bar F^2)
+ {i\over 4}(\alpha - \bar \alpha)\;[\;h^2 - (\partial \phi)^2 - 2 G^2 \;]
- {1\over 2}(\alpha +\bar \alpha)\;
\partial \phi \cdot h = 0~, \label{notophmod}
\ee
where
 $$ F^2 = F^{\alpha\beta}F_{\alpha\beta}~, \quad \bar F^2 =
F^{\dot\alpha\dot\beta}F_{\dot\alpha\dot\beta}
 $$
and $\alpha = \alpha(\phi)$. For the second solution (ii) in \p{sol12}
$\alpha =
\bar\alpha = -{1/\phi}$, so after the redefinition
\be
h^{\alpha\dot\alpha} \rightarrow \tilde{h}^{\alpha\dot\alpha} = \phi
h^{\alpha\dot\alpha} \label{redefh}
\ee
one gets the standard CS-term-modified Bianchi identity for
$\tilde{h}$
\be
\partial \cdot \tilde{h} - {i\over 4}\; (F^2 -
\bar F^2) = 0~. \label{notoph1}
\ee
It can still be solved through the antisymmetric gauge field (notoph) after
an appropriate shift of $\tilde{h}^{\alpha\dot\alpha}$ by the CS one-form.
This means that the solution (ii) in \p{sol12} corresponds just to the
non-linear  version of the VT multiplet discovered in \cite{dW1,dW2}.
At the same time, there is no way to reduce \p{notophmod} to
\p{notoph1} in the
new case corresponding to the solution (i) in \p{sol12}. There it is
impossible to solve the identity
\p{notophmod} through a notoph potential (at least, locally), though
we still end up with $8 + 8$ off-shell degrees of freedom. Thus we
encounter an essentially new version of the VT multiplet in this
case\footnote{The relation between these two non-linear versions of the VT
multiplet resembles that between the two well-known multiplets of $N=2$
supersymmetry without central charge, the tensor \cite{tenz} and non-linear
\cite{nonl} ones. Both of them have the same number of off-shell degrees of
freedom and in both cases there is a constraint on the vector component. In
the case of the tensor multiplet this constraint is of the notoph type
\p{notofbia} and it can be locally solved through the notoph potential. In
the case of the non-linear multiplet the constraint is modified and
resembles \p{notophmod} (it also contains terms bilinear in the vector
field strength in its r.h.s.). No local solution to this modified
constraint in terms of a gauge potential can be given.}.

It is easy to find the analogs of the free actions \p{act1}, \p{act2}
for both non-linear versions at hand. One starts from the Ansatz
\be
{\cal L}^{++} = A(L)\;D^+LD^+L + \bar A(L) \;\bar D^+L\bar D^+L
\label{ans}
\ee
and solves the differential equations for $A$, $\bar A$ following from
the analyticity constraint \p{analytL}. In both cases
(i), (ii) in \p{sol12} we get in this way two-parameter solutions for $A(L)$:
\bea
&& {\rm (i)}\; A(L) = d_1\; (\tan L + i) + d_2\;[1 + L(\tan L + i)]
\label{lagnov} \\
&& {\rm (ii)}\; A(L) = g_1\;{1\over L} + i g_2 \; L~, \label{lagBW}
\eea
where $d_{1,2}$, $g_{1,2}$ are arbitrary real constants.

The explicit form of the superfield Lagrangian density \p{ans}
in the most interesting case of the new solution (i) is
\be
{\cal L}^{++} = d_1\; (D^+)^2 L  + d_2\; [ D^+LD^+L + \bar D^+L\bar D^+L +
L(D^+)^2 L]~. \label{lagrnew}
\ee
It is instructive to work out the component
bosonic Lagrangian corresponding to \p{lagrnew}.  As a preparatory
step it is convenient to redefine $h^{\alpha\dot\alpha}$ as follows
\bea
&& h^{\alpha\dot\alpha} = {1\over \cos\phi}
\tilde{h}^{\alpha\dot\alpha}~, \label{redef1} \\
&& \partial \cdot \tilde{h} +{1\over 4}\;
(e^{i\phi}\; F^2 + e^{-i\phi}\; \bar F^2) +
{1\over 2 \cos\phi}\; \tilde{h}^2 - {1\over 2} \cos\phi
\; [(\partial \phi)^2 + 2 G^2 ] = 0~. \label{modbia}
\eea
Then a straightforward computation yields (up to an overall normalization
constant, modulo a total $x$-derivative and after putting
the auxiliary field $G=0$)
\be
{\cal L}_{bos} = v(\phi) \left\{ (\partial \phi)^2 -{1\over 2}
(F^2 + \bar F^2) -{i\over 2} (F^2 - \bar F^2)\;\tan\phi
-{1\over \cos^2\phi}\; \tilde{h}^2 -{2\over \cos\phi}
\;\partial \cdot \tilde{h} \right\}~, \label{bos}
\ee
where
$$
v(\phi) \equiv  d_1 + d_2 \phi~.
$$
Substituting \p{modbia} into this expression, we find the surprising
result
\be
{\cal L}_{bos} = 0\;\; \mbox{!}
\ee
Nevertheless, one can obtain a non-vanishing action after {\sl dualizing}
the notoph covariant field strength. This point is discussed in the next
Section.

As the last topic of this Section we present an alternative to the gauge
(\ref{gauge2}):
\be
\beta = b e^{i\psi}~, \quad b= {\rm const}\neq 0  \label{gauge1}
\ee
which yields
\be
(\gamma - \alpha)' = (\alpha - \gamma)\gamma - 2 b^2~, \quad
ib \psi' = b(\alpha - 2\gamma)~. \label{eqs1}
\ee
This time the analog of (\ref{selfcon5}) implies the additional constraint
\be
\alpha - \bar\beta - \gamma = 0~. \label{solfin}
\ee
Equations (\ref{eqs1}), (\ref{solfin}) have two different solutions:
\be
\alpha = b\;(2\; \cos 2\lambda + i \sin 2\lambda)~,
\;\; \gamma = b \;\cos 2\lambda~, \;\;
\beta = b\; e^{-2i\lambda}, \;\;
\lambda = \arctan a\; e^{-bL}, \label{sol1fin}
\ee
\be
\alpha = 2b~, \qquad \gamma = \beta = b~,  \label{sol2fin}
\ee
where $a,b$ are integration constants. The solution \p{sol2fin} is the $a=0$
contraction of the solution \p{sol1fin}, so the latter is more general.
We have
verified it to pass all conceivable self-consistency checks. Note that with
the choice of the $a=0$ version the constraints \p{nonlin1}, \p{nonlin2}
possess an important invariance under the shift $L \rightarrow L + {\rm
const} $, while it is not so in the general case $a\neq 0$. This invariance
guarantees the corresponding actions to be scale invariant, so the
parameter $a$ measures the breaking of such an invariance. Clearly, the
cases $a\neq 0$ and $a=0$ cannot be related by any field redefinitions,
since we have already fixed the reparametrization freedom while deriving
the above solutions.

The most general solution \p{sol1fin} was obtained in
the gauge \p{gauge1}, and it has the advantage of being non-singular in
the two important limits $a=0$ and $b=0$ which lead, respectively, to the
scale-invariant non-linear version \p{sol2fin} and to the linear
version. However, when constructing the invariant actions and inspecting
the deformations of the Bianchi identities in the general
$a \neq 0, b \neq 0$ case, it is more convenient to stay in the
gauge \p{gauge2}. The precise relation between the two gauges is as follows
\bea
(D^+)^2 \tilde L &=& (\bar D^+)^2 \tilde L = 2 c_1 b a \;
[ (\cot 2\lambda + i) D^+\tilde L D^+\tilde L
+ (\cot 2\lambda - i) \bar D^+\tilde L \bar D^+\tilde L ]~, \nn \\
D^+_\alpha
\bar D^+_{\dot\alpha} \tilde L &=& 0~, \nn \\
\tilde L &=& c_2  - {1\over 2c_1 b a} 2\lambda~. \label{rel12}
\eea
Here, $c_1$, $c_2$ are arbitrary integration constants reflecting the
residual freedom of shifting and rescaling $\tilde L$. They can always be
chosen so as to guarantee the limits $a=0$ and/or $b=0$ to be non-singular
in the gauge \p{gauge2} too. Finally, we note that it  is rather
straightforward to
check that in the case (ii) in (\ref{lagBW}) the invariants entering with
constants $g_1$ and $g_2$ take, respectively, the following form in the
gauge \p{gauge1}
\be
\sim e^{-bL}\;(D^+LD^+L + \bar D^+L\bar D^+L)~,  \label{nonlBW2}
\ee
and
\be
\sim i\;e^{-3bL}\;(D^+LD^+L - \bar D^+L\bar D^+L)~.  \label{nonlBW1}
\ee
Thus they generalize the Lagrangians ${\cal L}^{++}_2$ and ${\cal
L}^{++}_1$ of the linear case (eqs. \p{act2}, \p{act1}). Note that, as was
pointed out in the recent paper \cite{DK2}, the Lagrangian \p{nonlBW2} is a
total derivative like its linear version counterpart \p{act2}, and so it
gives rise to a topological invariant.

\setcounter{equation}{0}

\section{Dual versions of the VT actions}

The dual form of the above actions is obtained by implementing the notoph
constraint in the Lagrangian with the help of a
Lagrange multiplier. In the case of the constraint \p{modbia} this leads to
the action
\be
{\cal L}_{bos}{}' = - \lambda
\left( \partial \cdot \tilde{h} +{1\over 4}\;(e^{i\phi}\; F^2 + e^{-i\phi}\;
\bar F^2) +
{1\over 2 \cos\phi}\; \tilde{h}^2 - {1\over 2} \cos\phi
\;(\partial \phi)^2 \right)~. \label{duallagr}
\ee
Now $\tilde{h}^{\alpha\dot\alpha}$ is unconstrained, and one can
integrate it out by its algebraic equation of motion
\be
\tilde{h}^{\alpha\dot\alpha} = \cos\phi\; {\partial^{\alpha\dot\alpha}
\lambda \over \lambda}~. \label{elimin}
\ee
After that we get a typical sigma-model action
\be
{\cal L}_{bos}{}' =
-{\lambda \over 4}\;(e^{i\phi}\; F^2 + e^{-i\phi}\;
\bar F^2) + {\lambda \over 2}\; \cos\phi \;[(\partial \phi)^2 +
{(\partial \ln\lambda)^2}]~.
\label{dualbos}
\ee

Let us make once more an analogy with the non-linear $N=2$ multiplet. There
one cannot write down a non-vanishing (and $SU(2)$ invariant) action for
this multiplet itself \cite{harm,roc,IS}, but the dual action obtained by
implementing the defining constraint with the help of a Lagrange multiplier
yields a non-trivial sigma-model action in its bosonic sector.

No such subtleties occur in the case of the ``old'' non-linear version
corresponding to the solution (ii) \p{lagBW}. The only effect of
substituting the constraint \p{notoph1} into the appropriate analog of the
Lagrangian \p{bos} is the cancellation of the terms proportional to $g_1$,
in accord with the previous statement that the invariant proportional to $
g_1$ is a total derivative. In this case we have the following bosonic
Lagrangian (before dualization)
\be
{\cal L}_{bos} = g_2\; \left[\phi\; (\partial \phi)^2  -
{1\over 2}\;\phi\; (F^2 + \bar F^2) - {1\over \phi}\; \tilde{h}^2\right]~.
\label{bosstand}
\ee
The analog of the dual Lagrangian \p{dualbos} reads
\be
{\cal L}_{bos}{}' = \phi\; \left[g_2\; (\partial \phi)^2
+ {1\over 4g_2}\; (\partial \lambda)^2\right] -{1\over 2}(g_2\;\phi
+ {i\over
2}\lambda) \; F^2  -{1\over 2}(g_2\;\phi  - {i\over
2}\lambda) \; \bar F^2 ~.
\label{dualstand}
\ee

Both actions \p{dualbos} and \p{dualstand} can be recast in the generic
form of the bosonic part of the action of an $N=2$ gauge multiplet:
\be
{\cal L}_{bos}{}' = {i\over 2} (\partial{\cal F}'\; \partial\bar z  -
\partial z\; \partial\bar{\cal F}' + \bar{\cal F}''\;  F^2 -
{\cal F}''\;  \bar F^2 )~.\label{generic}
\ee
The holomorphic potential ${\cal F}(z)$ for the action  \p{dualbos} is
\be\label{newpot}
{\cal F}(z)={i\over 2} e^{-iz}\;, \qquad z=\phi + i\ln\lambda
\ee
and for the action  \p{dualstand}
\be \label{dewit}
{\cal F}(z)= -i{g_2\over 6}z^3\;, \qquad z= \phi - {i\over 2g_2}\lambda
 \; .
\ee
The potential (\ref{dewit}) can be obtained from that of ref. \cite{dW1},
\cite{dW2} by freezing the $N=2$ vector multiplet which gauges the central
charge. The potential (\ref{newpot}) is new and it would be interesting to
study whether it may occur in a stringy context.

The dualization procedure described above concerned the purely bosonic
sector of the action only. Carrying this procedure out in a fully off-shell
supersymmetric way is also possible. For simplicity here we explain this on
the example of the linear version of the VT multiplet. We take the
superspace action \p{action}, \p{act1} and add to it the harmonic
constraints \p{harm2a}, \p{harm2b} with analytic superfield Lagrange
multipliers:
\bea
S &=& \int d\zeta^{-4} \{[(\theta^+)^2 - (\bar\theta^+)^2](f^+f^+
- \bar f^+\bar
f^+)\nn \\&& + H^+D^{++}f^+ + \bar H^+D^{++}\bar f^+
 +\; G^{++}(D^{++}l + \theta^+f^+
+ \bar\theta^+\bar f^+ )\}\;.
\label{lm}
\eea
Note that the Lagrange multiplier $H^{+\alpha}$ has
a non-standard supersymmetry
transformation law in order to compensate for the variation of
the first term.
We assume that the central charge is still realized on
$f^+,\bar f^+$ as in
\p{5}, whereas on $l$ it acts as follows
\be\label{ccr}
\partial_5 l = {i\over 4} (\partial_{+\alpha}f^{+\alpha} -
\partial_{+\dot \alpha}\bar f^{+\dot \alpha})
\ee
(the reality condition \p{real} is not imposed at this stage, it appears
only as a result of the variation w.r.t. some Lagrange multiplier). This
realization of the central charge is compatible with supersymmetry. The
first term in \p{lm} is invariant under central charge transformations on
its own. The requirement of central charge invariance of the rest of the
action determines the central charge transformation properties of the
Lagrange multipliers:
\be
\partial_5 H^+_\alpha = -\partial_{\alpha\dot\alpha}\bar H^{+\dot\alpha} -
{i\over 4} \partial_{+\alpha} G^{++}\;, \quad \partial_5 G^{++}=0 \;.
\ee
To obtain the component content of the theory one should replace the
$\partial_5$-derivative terms contained in $D^{++}$ according to the above
rules and integrate over $\theta^+, \bar\theta^+$ and the harmonics. It can
be shown that upon elimination of the infinite set of auxiliary fields we
are left in the bosonic sector with two scalars and an abelian gauge vector
field, which belong to an on-shell $N=2$ vector multiplet dual to the
original VT one. More details and the treatment of the non-linear versions
will be given elsewhere.

\setcounter{equation}{0}

\section{Coupling to an external vector multiplet}

Here we shall deform the non-linear superfield constraints (their ``old''
version) to switch on the CS coupling to one external abelian vector
multiplet. The generalization to several such multiplets and to the
non-abelian case goes more or less straightforwardly and will be presented
elsewhere.

We choose the gauge \p{gauge1} and the simplest constant solution
to eqs. \p{eqs1}:
\be
\alpha = 2\gamma = 2 b, \quad \beta = be^{i\psi},
\quad b = {\rm const}~, \quad \psi = {\rm const}~. \label{simple}
\ee
Thus our starting point is the following set of constraints
\bea
(D^+)^2 L &=& 2 b D^+LD^+L + b e^{i\psi} \bar{D}^+L\bar{D}^+L \nn \\
D^+_\alpha \bar D^+_{\dot\alpha}L &=& b
D^+_\alpha L \bar D^+_{\dot\alpha} L  \label{flat}
\eea
As we saw before, this set corresponds to the non-linear version of
\cite{dW1,dW2} and it yields a CS-term modification of the Bianchi
identity for
the notoph field strength $h^{\alpha\dot\alpha}$. As we also saw, some
additional self-consistency conditions require for the given solution
\begin{equation}\label{111}
e^{i\psi} = 1
\end{equation}
at the level of the pure $L$ system. This phase factor
can be non-trivial in the presence of extra vector multiplets.

The abelian vector multiplet is represented by its superfield
strength $W$ which does not depend on the harmonics and obeys the chirality
condition and the Bianchi identity (reality condition)
\bea
&& D^{++} W = 0~, \nn \\
&& D^{\pm}_\alpha \bar W = \bar D^{\pm}_{\dot\alpha} W = 0, \nn \\
&& (D^+)^2 W =
(\bar D^+)^2 \bar W\;. \label{Wconstr}
\eea

In order to find an appropriate self-consistent deformation of \p{flat},
such that it is reduced to \p{flat} after switching off $W$, we proceed
in the following way. We start from the most general form of such
a deformation of the r.h.s. of eqs. \p{nonlin1}, \p{nonlin2}
consistent with the harmonic $U(1)$ charge $+2$ of the l.h.s. and
the constraint \p{harm2}. All the coefficients, including
$\alpha, \beta $ and $\gamma $, are assumed to be arbitrary functions
of $L$, $W$ and $\bar W$, with proper reality conditions imposed.
Next, just as in the pure $L$ case, we exploit the integrability
conditions  \p{selfcon}.
They lead to a huge number of differential and algebraic equations
on the coefficients. Among them we still have eqs. \p{sc1}, \p{sc2}.
To simplify the set of self-consistency conditions as much as
possible we utilize, like in the pure $L$ case, the reparametrization
freedom
\be
L \rightarrow L(\tilde{L}, W, \bar W)~. \label{reparLWW}
\ee
We can still impose the gauges \p{gauge1} or \p{gauge2} on the coefficients
$\alpha $, $\beta $, $\gamma $. We choose \p{gauge1}, with $b$ having no
dependence on $L$, $W$ and $\bar W$,
\be \label{const}
b = {\rm const}~.
\ee
There still remains the
freedom of shifting $L$ by a real function of $W$, $\bar W$. It can be used
to further restrict the r.h.s. of the deformed constraints.

Even after fixing the gauges we are still left with a considerable set of
equations. We first solve the equations \p{eqs1} for $\alpha $, $\beta $,
$\gamma $. As was stated above, for simplicity we choose the solution
\p{simple}, where $\psi $ is still independent of $L$ but now
depends on $W$, $\bar W$ (recall that $b = {\rm const}$ as a result
of fixing the gauge). This
dependence has to be specified by solving the rest of the consistency
conditions. Fortunately, the latter is greatly simplified under the choice
\p{const}.

As a result, we find the following most general self-consistent
deformation of the constraints \p{flat}
\bea
(D^+)^2 L &=& 2 b D^+LD^+L + b e^{i\psi} \bar{D}^+L\bar{D}^+L +
\lambda \bar{D}^+L\bar{D}^+\bar{W} +
\omega \bar D^+\bar W \bar D^+\bar W  \nonumber \\
&& + \;\nu (D^+)^2 W + \partial_W\nu D^+WD^+W +
\partial_{\bar W}\bar \nu e^{i\psi}\bar D^+\bar W \bar D^+ \bar W\;,
\label{def1a} \\
D^+_\alpha \bar D^+_{\dot\alpha}L &=& b D^+_\alpha L \bar D^+_{\dot\alpha} L
+ \sigma D^+_\alpha L \bar D^+_{\dot\alpha} \bar W  - \bar\sigma
\bar D^+_{\dot\alpha} L D^+_{\alpha} W - \omega D^+_\alpha W \bar
D^+_{\dot\alpha} \bar W \;.  \label{def2a}
\eea
Here all the coefficients, except for $\nu $, are expressed through
$\psi(W,\bar W)$
\be
\lambda  = i\;\partial_W \psi, \qquad \omega = -{1\over 4b}
\;\partial_W\psi\; \partial_{\bar W}\psi, \qquad
\sigma = {i\over 2}\;\partial_{\bar W}\psi, \label{expression}
\ee
Simultaneously, one gets the following remarkable equations for
$\psi $
\bea
\partial_W \psi &=& e^{i\psi} \partial_{\bar W}\psi, \label{phi1} \\
\partial_W\partial_{\bar W} \psi &=& 0 \;. \label{phi2}
\eea
The general solution of this system is given by
\be
e^{i\psi} = {1 + i\kappa \bar W\over 1 - i \kappa W}\;, \label{phi3}
\ee
or
\be
\psi = i\left( \ln(1-i\kappa W) - \ln(1+i\kappa \bar W)\right)
\;.
\label{phi4}
\ee
Here $\kappa$ is a real integration constant. We have also fixed one more
integration constant by requiring eq. \p{111} to be valid in the
pure $L$ limit $W = 0$ (or $\kappa = 0$).

Explicitly, the coefficients in \p{expression} are as follows
\be
\lambda = {i\kappa \over 1-i\kappa W}, \qquad
\sigma = {i\over 2}\;{\kappa \over 1 +i \kappa \bar W}, \qquad
\omega = - {1\over 4b}\;{\kappa^2 \over (1-i\kappa W)(1+i\kappa \bar W)}\;.
\label{explicit}
\ee
In order to get rid of the ``fake" singularity in $b$, one
should rescale
\be
\kappa = \sqrt{b}\tilde{\kappa}\;.
\ee

It remains to specify the coefficient $\nu $ in \p{def1a}. It
is given by the following expression
\be
\nu = \left[ e^{i\psi} a(W) + \bar a(\bar W) \right] e^{2bL}~, \label{nu}
\ee
$a(W)$ being an arbitrary holomorphic function.

The constraints \p{def1a}, \p{def2a} with the coefficients given by eqs.
\p{explicit} and \p{nu} describe the most general deformation of the `old'
nonlinear VT constraints \p{flat} in the presence of one extra vector
multiplet. It should be pointed out that the deformation presented here does
not distinguish an external vector multiplet from one that gauges
the central charge. Indeed, the above derivation relied merely upon
the anticommutativity of $D^+_\alpha,
\bar D^+_{\dot\alpha}$ and the constraints
\p{Wconstr}. These properties are valid irrespectively of whether $W$
is some external gauge superfield strength or it is the strength of
a superfield gauging the central charge.

An additional selection rule results from enforcing a self-consistency
condition like \p{selfcon5}. It leads to drastically different
consequences for the cases of rigid and gauged central charges \cite{DIKST}.
In the rigid case we are dealing with (when $W$ is treated as an external
$U(1)$ superfield gauge strength) it still requires
\p{111}~\footnote{Our special
thanks are due to S. Kuzenko for bringing up this point to us.}
\be \label{condrig}
e^{i\psi} = 1 \quad \Rightarrow \kappa = 0\;.
\ee
As a result, in this case the deformation above is fully specified by the
choice of the holomorphic function $a(W)$ in \p{nu}. The standard CS
modification of the Bianchi identity for $h^{\alpha\dot\alpha} $ arises for
$a(W) = c\; W$, $c$ being the appropriate CS coupling constant. However,
all the self-consistency conditions are still fulfilled by
an arbitrary $a(W)$.
Though the modified Bianchi identity has no local solution in
the general case, by analogy with the consideration in Sect. 4 we expect
that the `dualization' of this identity with the help of
a Lagrange multiplier vector multiplet may yield an acceptable
local theory.

As our last topic we give here the relevant invariant action.
The analytic Lagrangian density ${\cal L}^{++}$ for the $W$-deformed
case can
be constructed by the method of undetermined coefficients, like
we proceeded in
the previous Section. We compose the most general form of the Lagrange
density of charge $+2$
\bea
{\cal L}^{++} &=& g_1 D^+LD^+L + g_2\bar D^+L\bar D^+ L + g_3 D^+WD^+W
+g_4 \bar D^+\bar W\bar D^+ \bar W + g_5 D^+LD^+W  \nn \\
&& +\; g_6 \bar D^+L \bar D^+\bar W + g_7 (D^+)^2W~,
\eea
with all coefficients being arbitrary functions of $L, W, \bar W$.
Requiring ${\cal L}^{++}$ to be real imposes the following reality
conditions on $g_n$
\be
g_2 = \bar g_1, \;\; g_3 = \bar g_4, \;\;  g_5 = \bar g_6, \;\;
g_7 = \bar g_7\;. \label{reality}
\ee
Clearly, ${\cal L}^{++}$ obeys the condition
\be
D^{++}{\cal L}^{++} = 0\;.
\ee
Then we only need to extract the corollaries of the analyticity
constraint \p{analytL}. This requirement fixes ${\cal L}^{++}$ up to
three integration constants, thus yielding
three independent invariants
\bea
{\cal L}^{++}_{(1)} &\sim & e^{-bL} D^{+}LD^+L + {1\over b}\left(
e^{bL} -1 \right) (D^+)^2 G(W) + \mbox{c.c.}~, \label{L1} \\
{\cal L}^{++}_{(2)} &\sim & i \left\{ e^{-3bL} D^+LD^+L +
{2\over b} (D^+)^2 \left[\left(e^{-bL} - 1 \right) G(W) \right]
\right. \nonumber \\
&& \left. + \;{1\over b}\left( 1- e^{-bL} \right) (D^+)^2 G(W) -
\mbox{c.c.} \right\}~,
\label{L2} \\
{\cal L}^{++}_{(3)} &\sim & {1\over 2b} \left\{
\left( e^{-bL} - 1 \right) (D^+)^2 W + 2(D^+)^2\left[
\left(1- e^{-bL} \right) W\right] + \mbox{c.c.} \right\}~,
\label{L3}
\eea
where $G(W)$ is a ``potential'' for $a(W)$,
$$
a(W) =\partial_W G(W)~.
$$
The first two densities extend, respectively, the invariants
\p{nonlBW2} and
\p{nonlBW1}, while the third one is new, since it vanishes when $W=0$.
It still reduces to \p{nonlBW2} under the choice $W = const$. Note that
all these invariants were chosen to be well-defined in the limit
$b = 0$ by extracting some pure $W$ densities
\be
\sim (D^+)^2 {\cal F}(W) + (\bar D^+)^2 \bar{{\cal F}}(\bar W) \label{holom}
\ee
with some appropriate ${\cal F}$. They can be omitted without loss
of generality.

\vspace{0.3cm}

\noindent{\bf Acknowledgement.} We are indebted to B. de Wit, N. Dragon,
R. Grimm, M. Hasler, S. Kuzenko and B. Zupnik for valuable discussions. The
work of E.I. was partially supported by the grants of the Russian Foundation
for Basic Research RFBR 96-02-17634, RFBR-DFG 96-02-00180, by
INTAS grants INTAS-93-633, INTAS-94-2317 and by a grant of the Dutch NWO
Organization. He thanks D. L\"ust, C. Preitschopf and P. Sorba for their
kind hospitality at the Humboldt University, Berlin and at LAPP, Annecy,
where a part of this work was carried out.

\end{document}